\documentclass[pss,fleqn]{w-art}
\usepackage{times}
\usepackage{w-thm}

\usepackage[dvips]{graphicx}

\newcommand{\ybacuo}{$\rm YBa_2 Cu_3 O_{6+x}$}
\newcommand{\Tc}{$\rm T_c$}
\newcommand{\ybconi}{$\rm YBa_2(Cu_{1-y}Ni_{y})_3O_7$}

\newcommand{\ybco}{$\rm YBa_2 Cu_3 O_7$}

\newcommand{\BSCO} {$\rm Bi_2Sr_2CaCu_2O_{8+\delta}$}
\newcommand{\Tlbacuo} {$\rm Tl_2Ba_2CuO_{6+\delta}$}
\newcommand{\LSCO} {$\rm La_{2-x}Sr_{x}CuO_4$}

\begin{document}
\DOIsuffix{theDOIsuffix}
%%
%% issueinfo for header and copyright line
\Volume{XX}
\Issue{1}
\Copyrightissue{01}
\Month{01}
\Year{2004}

\pagespan{1}{}
%%
%%    Dates will be filled in by the publisher. The 'reviseddate' and
%%    'dateposted' (Published online) entry may be left empty.
\Receiveddate{\sf zzz} \Reviseddate{\sf zzz} \Accepteddate{\sf
zzz} \Dateposted{\sf zzz}
%%
%%    Give a maximum of six PACS code in numerical order.
%  [\it  maximum of 6 relevant codes, see \sf www.aip.org/pacs.]}
\title[Magnetic Resonant excitations in High-{$\rm T_c$} superconductors]
       {Magnetic resonant excitations in High-{$\rm T_c$} superconductors} 

\author[Y. Sidis {\it et al}]{Y. Sidis\inst{1}}
\author[]{S. Pailh\`es\inst{1}}
\author[]{ B. Keimer\inst{2}}
\author[]{P. Bourges\inst{1}}
\author[]{C. Ulrich\inst{2}}
\author[]{L.P. Regnault\inst{3}}

\address[\inst{1}]{Laboratoire L\'eon Brillouin, CEA-CNRS, CE-Saclay, 91191 Gif sur Yvette, France.}
\address[\inst{2}]{Max-Planck-Institut f\"ur Fertk\"orperforschung, 70569 Stuttgart, Germany.}
\address[\inst{3}]{CEA Grenoble, DRFMC  38054 Grenoble cedex 9, France.}
% D\'epartement de Recherche Fondamentale sur la Mati\`ere Condens\'ee,

\begin{abstract}
The observation of an unusual spin resonant excitation in the superconducting state of 
various High-\Tc ~copper oxides by inelastic neutron scattering measurements 
is reviewed. This magnetic mode % (that does not exist in conventional superconductors)
is discussed in light of a few theoretical models and likely corresponds to a 
spin-1 collective mode.
 \end{abstract}

\maketitle

%----------------------------------------------------------------------------
% \section{Introduction}

More than fifteen years after the high temperature superconductivity
discovery, antiferromagnetic (AF)
fluctuations pairing  mechanism\cite{afmodel} is still highly controversial. However, 
 inelastic neutron scattering (INS) measurements have successfully brought to light the existence 
of unusual AF excitations that develop below \Tc ~and could be the hallmark of an 
unexpected spin-1 collective mode, tightly bound to the superconducting (SC) state.
Whatever the role of that mode for superconductivity, it has to be derived from the 
same microscopic model used to discuss superconductivity. We here review its 
characteristic features in a few cuprates and discuss its possible origin in 
light of different theoretical models.

\begin{figure}[t]
\parbox{8 cm} {
\hspace*{-0.25 cm}
  \includegraphics[width=8cm]{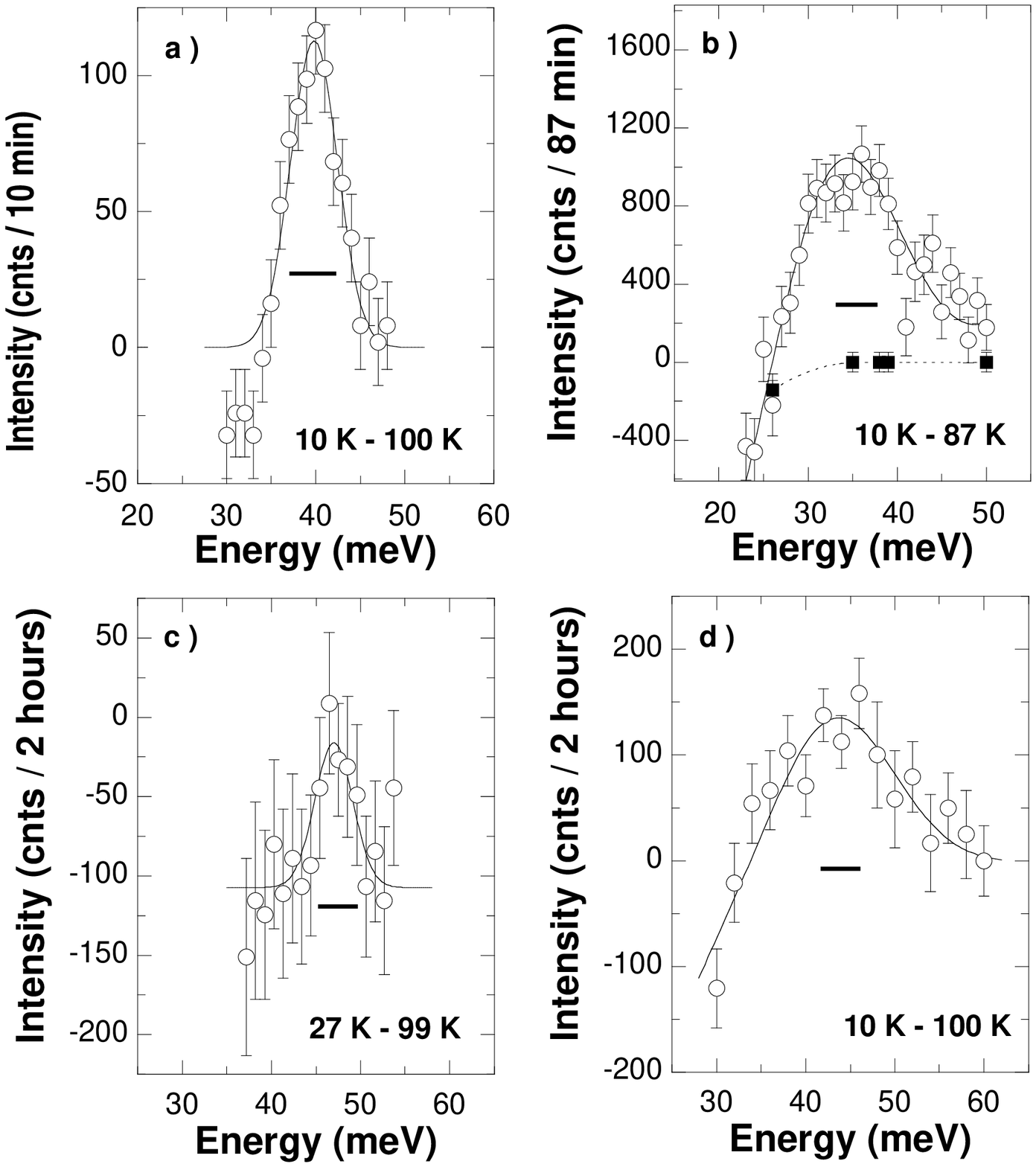}
% /home/llb8/sidis/user/text/SCNS-tex/res-scns.eps}
\caption[xxx]{Difference spectrum of the neutron intensities 
at low temperature, measured at the wave vector $\rm (\pi/a,\pi/a)$
and T$\ge$ $\rm T_c$:
a) $\rm YBa_2 Cu_3 O_{6.95}$: \Tc=93 K, V=10 cm$\rm ^{3}$ \cite{Fong95},
b) $\rm YBa_2(Cu_{1-y}Ni_{y})_3O_7$: \Tc=80 K,  V$\sim$2 cm$\rm ^{3}$ \cite{Sidis00},
c) $\rm Tl_2Ba_2CuO_{6+\delta}$:\Tc$\sim$90 K,  V=0.11 cm$\rm ^{3}$ \cite{He-Science},
d) $\rm Bi_2Sr_2CaCu_2O_{8+\delta}$; \Tc=91 K,  V=0.06 cm$\rm ^{3}$ \cite{Bi2212}.
Data are fitted to a Gaussian profile. The solid bar indicates the energy resolution. 
\label{fig-resonance} } }
\hspace*{+.5 cm}\parbox{6 cm} {
% \hspace*{+.5 cm}
\includegraphics[width=6.2cm,height=7.3 cm]{lt-all2.epsi}
% /home/llb8/sidis/user/text/SCNS-tex/lt-all2.epsi}
\caption[xxx]{ Doping dependence of the energy of the magnetic resonance peak at
$\rm (\pi/a,\pi/a)$, $\rm E_r$, and of twice the maximum of the superconducting gap, 
2 $\rm \Delta _m$. $\rm E_r$ has been measured by INS  in YBCO 
 \cite{Rossat91,Mook93,Fong95,Bourges96,Fong2000,dai2001}, and
BSCO\cite{Bi2212,He2000}. 2 $\rm \Delta _m$ is determined from angle resolved photo-emission 
spectroscopy (ARPES) \cite{Mesot99} or superconducting-insulator-superconductor
(SIS) tunneling \cite{Zasadzinski01} data performed in BSCO.
The doping level is estimated through the empirical relation: 
\Tc/\Tc$\rm ^{max}$=1-82.6($\rm n_h$-$\rm n_{opt}$)$\rm ^{2}$ \cite{Tallon95}.
\label{figErvsdoping} } }
\end{figure}

% \section{INS experiments}
In optimally doped \ybacuo ~(YBCO) (\Tc=93 K) where it has been discovered \cite{Rossat91} 
(Fig.~\ref{fig-resonance}.a), the spin excitation  spectrum is dominated in the SC 
state by a sharp magnetic excitation at an energy of $\sim$40 meV and at the  planar 
antiferromagnetic wave vector 
$\rm {\bf q}_{AF}=\rm (\pi/a,\pi/ a)$, the so-called magnetic resonance 
peak \cite{Rossat91,Mook93,Fong95,Bourges96}.
% Regnault94,Fong95,Bourges96,Regnault98}.
Its intensity decreases with increasing temperature and vanishes steeply at
\Tc, without any significant shift of its characteristic energy
$\rm E_r$. In the underdoped regime,  $\rm E_r$ monotonically decreases  with decreasing 
hole concentration \cite{Fong2000,dai2001}
 % \cite{Dai96,Fong97,Bourges97,Fong2000}
 so that $\rm E_r \simeq$ 
5 $\rm k_B T_c$ (Fig.~\ref{figErvsdoping}). Besides, it is possible to vary
\Tc ~without changing the carrier concentration through impurity substitutions of Cu 
in the $\rm CuO_2$ planes. This is the case in \ybconi ~(y=1\%, \Tc=80 K), where the 
magnetic resonance peak shifts to lower energy with a preserved $\rm E_r$/$\rm k_B T_c$ 
ratio (Fig.~\ref{fig-resonance}.b) \cite{Sidis00}.

In optimally doped \BSCO ~(BSCO) (\Tc=91 K), a similar magnetic resonance peak has been observed 
at 43 meV  (Fig.~\ref{fig-resonance}.d) \cite{Bi2212}. Furthermore, $\rm E_r$ shifts  
down to 38 meV in the overdoped regime ($\rm T_c$=80 K) \cite{He2000}, preserving a 
constant ratio with \Tc: $\rm E_r  \simeq$  5.4 $\rm k_B T_c$ (Fig.~\ref{figErvsdoping}).
Thus, whatever the hole doping, the energy position of the magnetic resonance peak always 
scales with $\rm T_c$. In contrast to YBCO, where the resonance
peak is resolution limited in energy, the resonance peak in BSCO exhibits an
energy width of $\sim$13 meV. In addition, the momentum width of the excitation
is twice broader . A similar energy and momentum broadening has been also reported in 
\ybconi ~\cite{Sidis00} (Fig.~\ref{fig-resonance}.b) and  can therefore be ascribed to  
disorder, such as impurities or inhomogeneities.
Furthermore, the observation of a spatial distribution of the SC gap in \BSCO ~
by Scanning Tunneling Microscopy measurements \cite{STM} provides evidence in favor of 
an intrinsic disorder in this system.

Further, the magnetic resonance peak has been observed in optimally doped
\Tlbacuo (\Tc=90 K) at $\rm E_r$=47 meV  (Fig.~\ref{fig-resonance}.c) \cite{He-Science}. 
This yields a ratio $\rm E_r$/$\rm k_B T_c \simeq$ 6 slightly larger than \ybacuo. 
Nevertheless, as in \ybco, the
excitation is limited by the resolution in energy and displays a momentum
width of 0.25 \AA$^{-1}$ (half width at half maximum).
Meanwhile, the energy integrated intensity of the magnetic resonance peak is
almost the same in both systems: 0.7-0.8 $\rm \mu_B^2.eV^{-1}/CuO_2 $ plane. 
Thus, the magnetic
resonance peak appears as a common excitation to the SC state of 
all High-\Tc ~superconductors, investigated so far by INS measurements, whose maximum 
\Tc ~can be as high as $\sim$90 K. Furthermore, the existence of this
excitation does not depend on the number of $\rm CuO_2$ planes per
unit cell: one for \Tlbacuo ~and two for \ybacuo ~and \BSCO.

While the magnetic resonance peak exists in cuprates whose maximum \Tc ~is
about 90 K, it has never been observed in the  mono-layer system \LSCO
~with a maximum of \Tc ~of $\sim$ 40 K. Furthermore, the magnetic 
excitations in that compound are
rather strong even in the normal state and located at incommensurate
planar wave vector $Q_{mag} = \rm (\pi/a (1\pm \delta_{inc}), \pi /a)$ and  
$\rm (\pi / a, \pi /a (1\pm \delta_{inc}))$ \cite{aeppli}. This is in a marked contrast 
with the the systems mentioned above, for which the normal state magnetic fluctuations 
(if observable) remain centered around $\rm (\pi / a, \pi / a)$. However, passing 
through \Tc, the incommensurate spin fluctuations of \LSCO ~are enhanced and become
narrower in momentum space, in an energy range which is about 5$\rm k_B$\Tc ~\cite{Mason96}.
This phenomenon, usually referred to as a "coherence effect" and the resonance
peak  could eventually share a common origin. 

\begin{figure}[t]
\parbox{7 cm} {
\hspace*{-0.25 cm}
  \includegraphics[width=8 cm]{pbop.epsi}
  \caption[xxx]{a) Dispersion of the magnetic resonance peak as a function 
q=$\rm (\pi /a (1\pm \delta_{inc}), \pi /a)$, measured in $\rm YBa_2 Cu_3 O_{6.85}$ 
\cite{Bourges-science}, b) dispersion of the spin exciton \cite{Onufrieva02} (thick line)
below the continuum which is represented by the dashed area. The maximum of the 
dashed lines correspond to $2 \Delta_m$, and their crossing defines 
 $\omega_c=2 \Delta_{k_s}$.
\label{fig1} } }
\hspace*{+.5 cm}\parbox{7.5 cm} {
 \hspace*{+1. cm}\includegraphics[width=6.5cm,height=4.8 cm]{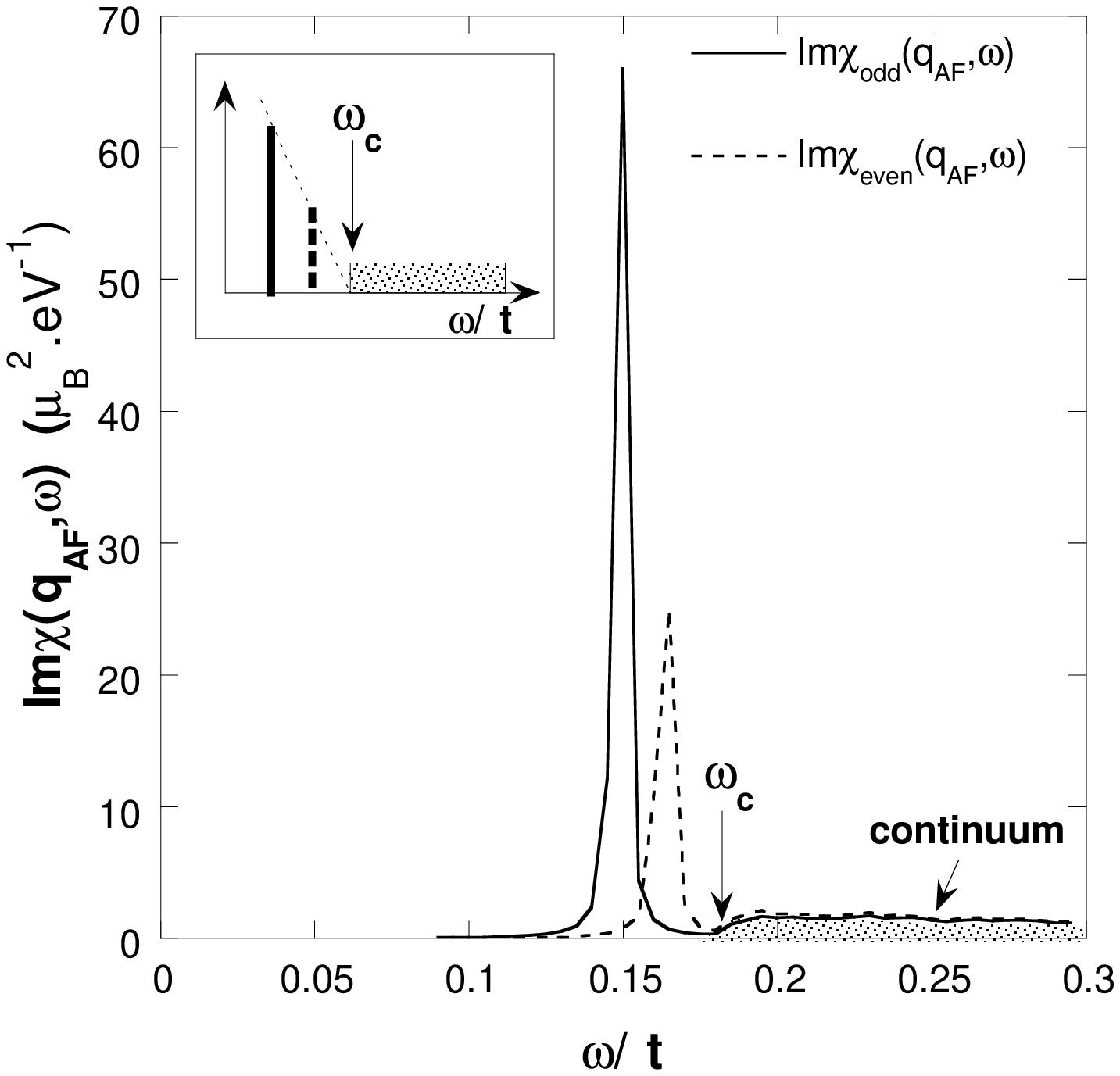}
\caption[xxx]{ Calculated spin susceptibility for both odd and even symmetry 
according to Eq. \ref{eq2} with an interlayer coupling of $J_{\perp} \simeq$ 0.1 J
and with the band structure parameters and $t/J$=2 of ref. \cite{Onufrieva02} and
$t$=250 meV. 
%($t$=250 meV, $J$=125 meV). 
The dotted areas show the weak electron-hole continuum intensity.
\label{theo-chi} } }
\end{figure}

In underdoped \ybacuo ~(x=0.6,$\rm T_c$=63 K, $\rm E_r$=34 meV), INS measurements 
provide evidence for incommensurate-like spin fluctuations at 24 meV and low temperature 
(seemingly similar to those observed \LSCO) \cite{Mook-incom}.
These incommensurate-like spin fluctuations are also observed at higher oxygen 
concentrations: x=0.7\cite{Arai}, x=0.85 \cite{Bourges-science}. 
As a function of temperature and energy 
\cite{Bourges-science}, the incommensurability ($\rm \delta_{inc}$) increases 
below $\rm T_c$ with decreasing temperature and decreases upon approaching 
$\rm E_r$ in the SC state (Fig.~\ref{fig1}.a). The simultaneous disappearance of 
$\rm \delta_{inc}$ at $\rm E_r$ and \Tc ~ indicates that the resonance peak and the 
incommensurate-like spin fluctuations are intrinsic features of the SC state and 
that they can be viewed as continuously connected (Fig.~\ref{fig1}.a). In other 
words, these results lead to an unified description of both the incommensurate 
spin excitations and the magnetic resonance peak in terms of a unique collective 
spin excitation mode with a downward dispersion\cite{Bourges-science}. Recently, 
the actual symmetry of this dispersion for an optimally doped YBCO sample 
has been looked at carefully\cite{dmitry} and was found basically circular 
within the 2D copper-oxygen plane. In addition, a second magnetic mode with much
weaker intensity is reported dispersing upward above the 
$\rm (\pi/a,\pi/a)$ peak\cite{dmitry}.  The deep underdoped 
state YBCO$_{6.5}$ has been also recently re-investigated in partly detwinned 
sample with an ortho-II structure\cite{buyers}. In contrast to
the dispersive mode picture, it is  claimed\cite{buyers} that the low 
energy magnetic excitations are essentially one-dimensional as expected 
for  hydrodynamic stripes. Therefore, the detailed doping 
dependence of the spin fluctuations needs to be clarified to reconcile these 
conclusions by studying fully detwinned samples. 
Indeed, one needs to determine the specific role of the Cu-O chains 
in YBCO for the magnetic anisotropy. 

The main difference between mono-layer and bilayer systems shows up in the
momentum dependence of the magnetic resonance peak along the (0 0 1) direction. 
In \Tlbacuo, the excitation remains purely bidimensional. In contrast in bilayer systems,
the two $\rm CuO_2$ planes correlate antiferromagnetically within the bilayer.
That interlayer AF coupling is responsible in insulating parent compounds for both 
acoustic and optic magnons, whose counterpart in the metallic state are the odd (o) 
and even (e) excitations. The neutron scattering cross section  then 
reads\cite{Fong2000}:

\begin{equation}
\frac{ d^2 \sigma (Q,\omega)}{d \Omega d  \omega} \propto 
\sin^2(Q_z d/2) Im \lbrack \chi_{o}(Q,\omega) \rbrack  +
\cos^2(Q_z d/2) Im \lbrack \chi_{e}(Q,\omega) \rbrack\label{eq-bilayer}
\end{equation}

where $\rm  Im \lbrack \chi_{o,e}(Q,\omega) \rbrack$ corresponds to the imaginary part 
of the dynamical magnetic susceptibility in each channel and $d$   (=3.3 \AA) stands 
for the distance between $\rm CuO_2$ planes within the bilayer.  Intuitively, one 
could expect a splitting of the magnetic resonance peak under the interlayer AF coupling, 
leading to a magnetic resonance in each channel. For a long time, the magnetic resonance 
peak was observed only in the odd channel. However, we could recently observe a
resonant mode in each channel in slightly overdoped YBCO through 10 \% substitution of Y by Ca
\cite{pailhes}. They occur at two different energies $E_r^o$ = 36 meV and 
$E_r^e$ = 43 meV and the even mode exhibits an intensity one third times less than 
the odd one. The question why the even mode is now sizeable in this overdoped regime 
and not in previous studies remains open as it could be simply due to 
improvement of neutron instruments. However, it might also be related to 
the electronic transport between closely spaced $\rm CuO_2$ layers which becomes 
coherent in the overdoped regime, as demonstrated by recent
experiments showing well-defined bonding and antibonding bands.

%----------------------------------------------------------------------------
% \section{Spin 1-collective mode}

Considering the different models for the magnetic resonance peak, 
we focus on models where electron-electron interactions play the 
central role, despite the still possible existence of electron-phonon 
couplings in cuprates. Essentially, there is no indication
of an effect of the lattice on the magnetic resonance peak. Secondly, 
we consider here models where the downward dispersion of the resonant mode
would naturally emerge. For instance, approaches \cite{morr,sachdev} which 
associate the resonance peak to a pre-existing soft mode reminiscent of 
nearby (commensurate or incommensurate) AF phase would yield a collective 
mode dispersing predominantly upward.

The existence of a spin 1-collective in d-wave superconductors such as 
High-\Tc ~ cuprates, is derived from an itinerant description of the 
magnetic properties of the system and of strong correlation effects
(\cite{Liu95,Millis96,Abanov99,Norman00,eremin,Onufrieva02} and references therein).
This leads to a particle-hole ({\it p-h}) bound state, usually referred as a spin 
exciton. In these strong coupling models (see also Refs. \cite{bl,sega} in 
the framework of the $t-J$ model),  
the generalized spin susceptibility $\rm \chi(q,\omega)$ is expressed as a function of the
non interaction spin susceptibility $\rm \chi_0(q,\omega)$ and the
magnetic interaction. $\rm \chi(q,\omega)$ has an RPA-like form:

\begin{equation}
\chi(q,\omega)=\frac{\chi_0(q,\omega)}{1+J(q)\chi_0(q,\omega)}
\label{eq1}
\end{equation}
where $\rm  J(q)=2J(cos(q_x)+cos(q_y))$ is the intra-plane
AF super-exchange coupling and $\rm \chi_0(q,\omega)$ describes in an itinerant system 
the continuum of spin flip particle-hole ({\em p-h}) excitations, given by the Lindhard
function in the normal state or the BCS function in the SC state. In the SC state, due 
to the opening of the SC gap, the continuum becomes gaped (Fig.~\ref{fig1}.b) below
a threshold energy at $\omega_c= 2\Delta_{k_s}$,
where $\Delta_{k}$ is the momentum dependent superconducting $d$-wave energy gap
and $k_s$ is the so-called hot-spot wave vector defined as both 
$k_s$  and $k_s+Q_{AF}$ are lying on the Fermi surface. 
In addition to the {\em p-h} excitations within the continuum, a  spin triplet {\em p-h}
bound state can form  below the continuum thank to the AF interaction.
In Eq.~\ref{eq1}, the dynamical Stoner criterion, i.e. 
$1+J(q)Re \lbrack \chi_0(q,\omega) \rbrack$=0, is then fulfilled for an energy smaller 
that the threshold of the continuum at wave vector q.
This spin 1-collective  mode is characterized
by a downward dispersion controlled by the momentum dependences of the continuum 
threshold and the magnetic interaction (Fig~\ref{fig1}.b). The mode vanishes when 
approaching the continuum by changing wave vector from $\rm (\pi/a,\pi/a)$. 
The model also predicted an excitation dispersing upward within the continuum  
\cite{Norman00,Onufrieva02} which might correspond to the recently observed 
high energy mode\cite{dmitry}.

From experimental point of view, a major challenge for future INS experiments 
would be to observe the magnetic continuum. So far, one can deduce
the continuum threshold, $\rm \omega_c$, from i) the measurement of the maximum 
of the SC gap, $\rm \Delta_m$ (determined by angle resolved photo-emission 
spectroscopy\cite{Mesot99}, by the measurement of the $\rm B_{1g}$ mode in Raman 
scattering\cite{raman} or by tunneling data\cite{Zasadzinski01}, 
see Fig. \ref{figErvsdoping}) and ii) from the Fermi surface 
topology\cite{fermisurface}. At optimal doping in BSCO, one can show that the 
magnetic resonance peak lies well below the continuum:  $\rm \Delta_m \simeq$ 
35 meV, $\rm \omega_c \simeq$ 1.8$\rm \Delta_m$ 
and $\rm E_r \simeq$ 1.2$\rm \Delta_m$. %  \cite{Bourges-LT23}.
From optimal doping to the overdoped regime, one may expect $\rm \omega_c$ to 
reach the limit $\sim$2$\rm \Delta _m$. Simultaneously, the ratio 
2$\rm \Delta _m$/$\rm k_B T_c$ 
evolves from 7-8 to 5-6, while according to INS data, the ratio $\rm E_r$/$\rm k_B T_c$ 
seems to be preserved (Fig.~\ref{figErvsdoping}). These evolutions suggest that in the 
overdoped regime the binding energy of the spin exciton, $\rm  \omega_c$-$\rm E_r$,
weakens, leading to the possible disappearance of the spin 1-collective mode.
Further INS experiments, in the deeply overdoped regime are required to test
such a possibility.

For a bilayer system \cite{Liu95,Millis96,Li99}, the interlayer AF coupling, 
$\rm J_{\perp}$, is treated in perturbation, 
such that the odd (o) and even (e) spin susceptilities are given by:
\begin{equation}
\chi_{o}(q,\omega)=\frac{\chi(q,\omega)}{1 - J_{\perp}\chi(q,\omega)} \ \  {\rm  and }  \ \  
\chi_{e}(q,\omega)=\frac{\chi(q,\omega)}{1 + J_{\perp}\chi(q,\omega)}
\label{eq2}.
\end{equation}

In Fig. \ref{theo-chi}, we calculated these susceptibilities with a realistic value of 
$\rm J_{\perp}$ ($\sim$ 0.1 J) \cite{Reznik96} and with other parameters identical to
those of ref. \cite{Onufrieva02}. Because of $\rm J_{\perp}$, the odd 
spin exciton is likely pushed to lower energy, while the even one merges 
into the continuum. Actually, this explains why the odd channel mode is mainly observed
as well as why the resonant mode appears to shift to lower energy in a bilayer system
in regard with a  mono-layer system. Going further, their respective spectral weights
$W_{o,e}$ are predicted to be approximately proportional to their binding
energies \cite{Millis96} as $W_{o,e} \sim \omega_c-E_r^{o,e}$. This  is found 
in the calculated susceptibilities in Fig. \ref{theo-chi} and sketched in the 
figure inset. Using this property in overdoped YBCO-Ca\cite{pailhes}, one could
directly estimate of continuum threshold at $\omega_c(Q_{AF}) \simeq 49$ meV.

If the spin exciton scenario provides a plausible explanation for the resonance 
peak around the optimal doping, more theoretical work is needed to fully account 
for its evolution as a function of hole doping. In particular, there is no 
explicit relationship between the
energy of the collective mode at $\rm (\pi /a, \pi/a)$ and the value of \Tc.
The phenomenological relationship $\rm E_r$/$\rm k_B T_c \simeq$ 5-6 
remains to be explained. Furthermore, the same model must simultaneously
describe the unusual features observed below \Tc ~ as well as the spin excitation
spectrum of the normal state. In the underdoped regime, the magnitude and the energy 
dependence of spin fluctuations observed by INS are still difficult to reproduce 
quantitatively. Thus, most likely, the model needs to go beyond a purely itinerant 
picture in order to capture the deep underdoped 
state.

However the spin-exciton interpretation for the magnetic neutron resonance 
and its downward dispersion
is not unique. Indeed, besides such an approach the spin 1-collective mode 
corresponds to a {\it p-h} bound state, it has also been proposed that the 
magnetic resonance peak could be a {\it p-p} pair, the $\pi$-excitation, in 
SO(5) model (spin itinerant picture) or a magnon-like excitation 
(spin localized picture) of a disordered stripe phase.

The SO(5) model \cite{Zhang97} considers as a starting point that, in the AF 
insulating state of  cuprates, there exists a super-symmetry that allows the system to 
switch from the AF state to the d-wave SC state. This symmetry
involves the existence of a Goldstone mode, the $\pi$-excitation \cite{Demler98}. 
This excitation can be depicted as an excitation that creates a {\it p-p} pair carrying 
a charge 2e, a spin S=1 and with total momentum $\rm(\pi/a ,\pi/a)$. Upon
doping the symmetry is broken, the $\pi$-excitation survives, but becomes massive.
While it exists already in the normal state, it can only be observed by INS in the 
SC state due to the {\it p-h} mixing. Its characteristic energy is roughly linear
with hole doping and its intensity in the SC state scales with $\rm | \Delta_m |^2$. 
Moreover, a downward  dispersion is obtained due to by the phase slips of the SC 
order parameter induced by the propagation of the $\pi$-excitation\cite{Hu01}. 
Such an approach accounts for several features observed by INS, especially in the 
underdoped regime, where \Tc ~ and $\rm E_r$ increase with hole doping. 
% Through Zn substitution, the hole doping is unchanged, while the SC condensate is reduced.
The $\pi$-excitation should decrease in magnitude, but remains almost at the same 
energy, as observed experimentally. However this scenario cannot 
explain the decrease of $\rm E_r$ in the overdoped regime. Furthermore, recent 
calculations have shown that, if the $\pi$-excitation existed, it should be observed 
at high energy, above 2$\rm \Delta _m$ \cite{Chubukov-pi}: this casts some doubt 
about the interpretation of the resonance as a $\pi$-triplet excitation.

Alternatively, the stripe model  %(see Ref.~\cite{Emery-review} and references therein) 
considers that in a S=1/2 AF Heisenberg system, doped holes
segregates to form lines of charges, separating AF domains in anti-phase. 
The metallic state is viewed as a disordered stripe phase, where charged
lines can fluctuate. While there is not a general interpretation of the
magnetic resonance peak in stripe models, it has been for instance proposed
that the resonance peak and the incommensurate spin fluctuations
observed by INS in the SC state could be viewed  as magnon-like excitations 
reminiscent of the ordered stripe phase\cite{Batista,kruger} or could 
correspond to the eigen  magnetic modes of the liquid stripe phase 
\cite{Morais-smith}. Magnons, developing symmetrically around the magnetic 
incommensurate wave-vector, $Q_{mag}$, of the stripe ordered phase and 
merging at $\rm (\pi/a,\pi/a)$, actually describe correctly the spin dynamics 
observed in stripe-ordered nickelates\cite{lsno} as predicted in the spin-only 
model\cite{Batista,kruger}. In cuprates, the lack of symmetric peaks around 
the incommensurate wave-vector
$Q_{mag}$ (see Fig. \ref{fig1}) does not seem to validate these approaches. 
In addition, this model, if interesting, fails to explain why the resonance peak 
and the incommensurate spin fluctuations exist basically only in the SC state. 
Independently, it could be also interesting to understand how, in bilayer compounds, 
the adjacent ${\rm CuO_2}$ planes succeed in accommodating the Coulomb repulsion 
between charged lines and the AF interlayer coupling. This is a central issue to 
account for as the magnetic resonance peak exists mostly in the odd channel.

% \section{Conclusion}

Finally, INS experiments have shown the existence of an usual enhancement of spin fluctuations
in the SC state around the vector $\rm (\pi/a, \pi/a)$ and at an energy
$\rm E_r$ which is found experimentally to scale with \Tc. Combined with the 
observation at lower energy of incommensurate spin fluctuations, that develop also 
below \Tc, INS data point toward the existence of a dispersive spin 1-collective mode 
deep inside the SC state. The observation of that mode, first discovered in \ybco, has 
been then extended to other systems with one or two $\rm CuO_2$ planes per unit cell, 
such as \BSCO ~and \Tlbacuo. This establishes the
magnetic resonance peak as a generic excitation of the SC state of cuprates
whose maximum \Tc ~can be as high as 90 K.
In the strongly under- and overdoped regimes (\Tc ~$\le$ 50 K) or
in other cuprate families with lower maximum \Tc  (such as \LSCO), the observation of the
the spin 1-collective mode (if any) still requires more experimental work.
% To conclude, whatever the correct interpretation for the magnetic resonance peak, 
In any case, the 
observation of such an excitation, thank to inelastic neutron scattering, is one of 
the most persuasive experimental indication
of the crucial role of magnetic interactions in for the physics of High-\Tc ~copper oxides.


\begin{thebibliography}{10}

\bibitem{afmodel} See e.g. D. Scalapino, Phys. Rep. {\bf 250} 329 (1995).

% ----------------------------------resonance YBCO7

\bibitem{Rossat91}  J. Rossat-Mignod, L.P. Regnault, C. Vettier, P. Bourges, P. Burlet, J. Bossy, {\em et al.}, Physica C {\bf 185-189},
86 (1991).

 % J. Rossat-Mignod, L.P. Regnault, C. Vettier, Ph. Bourges, P. Burlet, J. Bossy, 
 % J.Y. Henry and G. Lapertot : Physica C, {\bf 185-189}, 86-92 (1991).
 
\bibitem{Mook93}  H.A. Mook, M. Yethiraj, G. Aeppli, T.E. Mason, and T. Armstrong, Phys. Rev. Lett. {\bf 70}, 3490
(1993).

% \bibitem{Regnault94}  L.P. Regnault {\em et al.},  Physica C {\bf 235-240},
% 59 (1994); Physica B {\bf 213-214}, 48 (1994).

\bibitem{Fong95}  H. F. Fong, B. Keimer, P.W. Anderson, {\em et al.},  Phys. Rev. Lett. {\bf 75}, 316 (1995); Phys. Rev. B {\bf 54}, 6708 (1996).

 % Phys. Rev. Lett. {\bf 75}, 316
\bibitem{Bourges96} P. Bourges, L.P. Regnault, Y. Sidis, and C. Vettier, Phys. Rev. B {\bf 53}, 876
(1996).

% \bibitem{Regnault98}  L.P. Regnault {\em et al.},  in 
% {\it Neutron Scattering in Layered Copper-Oxide Superconductors}, A. Furrer {\it et al.} 
% eds (Kluwer, Amsterdam, 1998). 

% ----------------------------------Underdoped YBCO


% \bibitem{Dai96}  P. Dai {\em et al.}, Phys. Rev. Lett. {\bf 77}, 5425 (1996).

% \bibitem{Fong97}  H.F. Fong {\em et al.},  Phys. Rev. Lett. {\bf 78}, 713 (1997).
 
% \bibitem{Bourges97}  P. Bourges {\em et al.}, Europhys. Lett. {\bf 38}, 313 (1997).

\bibitem{Fong2000}  H.F. Fong, P. Bourges, Y. Sidis, L.P. Regnault, J. Bossy,  A.S. Ivanov, {\em et al.}, 
Phys. Rev. B {\bf 61}, 14774 (2000).

\bibitem{dai2001} % P. Dai {\it et al.}, 
P. Dai, H.A. Mook,R. D. Hunt, F. Do\u{g}an, Phys. Rev B., {\bf 63}, 054525 (2001). 
% ----------------------------------Compare Ni vs ZN

\bibitem{Sidis00} %  Y. Sidis {\em et al.},
 Y. Sidis, P. Bourges, H. F. Fong, B. Keimer, L. P. Regnault, J. Bossy, {\em et al.}, Phys. Rev. Lett. {\bf 86}, 4100 (2001).

% ----------------------------------BISCO


\bibitem{Bi2212}  %  H.F. Fong {\em et al.}, 
H.F. Fong, P. Bourges, Y. Sidis, L.P. Regnault, A.S. Ivanov,
G.D. Gu, {\em et al.}, Nature {\bf 398}, 588 (1999).

\bibitem{He2000}  % H. He {\em et al.},
 H. He, Y. Sidis, Ph. Bourges, G.D. Gu, A.Ivanov, N. Koshizuka, 
       B. Liang, {\em et al.}, Phys. Rev. Lett. {\bf 86}, 1610 (2001).

\bibitem{STM} % K.M. Lang {\em et al.}, 
 K.M. Lang, V. Madhavan, J.E. Hoffman, E. W. Hudson, H. Eisaki, S. Uchida, J.C. Davis, Nature {\bf 415}, 412 (2002).


%----------------------------------TlBaCuO

\bibitem{He-Science} H.He, P. Bourges, Y. Sidis, C. Ulrich
L.P. Regnault, S. Pailh\`es, {\em et al.}, Science {\bf 295}, 1045 (2002).

%-----------------------------------Er=5.3 KBTc
\bibitem{Mesot99} J. Mesot, M. R. Norman, H. Ding, M. Randeria, J. C. Campuzano, {\em et al.}, Phys. Rev. Lett. {\bf 83}, 840 (1999).

\bibitem{Zasadzinski01} J.F. Zasadzinski, L. Ozyuzer, N. Miyakawa,  {\em et al.}, 
% K.E. Gray, D.G. Hinks, C. Kendziora,
% J.F. Zasadzinski {\em et al.}, 
{\rm Phys. Rev. Lett.} {\bf 87}, 067005 (2001)  (cond-mat/0102475).

\bibitem{Tallon95} J.L. Tallon, C. Bernhard, H. Shaked, R. L. Hitterman and J. D. Jorgensen, Phys. Rev. B {\bf 51}, 12911 (1995).


%----------------------------------LSCO

% \bibitem{Mason98} T.E. Mason, in {\em Neutron Scattering Studies of Spin Fluctuations in High Temperature Superconductor}, Handbook on the Physics and Chemistry of Rare Earths, Special Volumes on High Temperature Rare Earth Superconductors, Eds K.A. Gschneider Jr., L. Eyring and K.B. Maple (1998).

\bibitem{aeppli}  % G. Aeppli, {\it et al.}, 
G. Aeppli, T. E. Mason, S. M. Hayden, H. A. Mook, J. Kulda,  { Science} {\bf 278}, 1432 (1997).

\bibitem{Mason96} % T.E. Mason {\em et al.},
 T. E. Mason, A. Schr\"oder, G. Aeppli, H. A. Mook, and S. M. Hayden,
  Phys. Rev. Lett {\bf 77}, 1604 (1996).


% ---------------------------------- YBCO- incom


% \bibitem{Dai-incom}  P. Dai {\em et al.}, Phys. Rev. Lett. {\bf 80}, 1738 (1998).

\bibitem{Mook-incom} % H.A. Mook {\em et al.}, 
H. A. Mook, P. Dai, S. M. Hayden, G. Aeppli, T. G. Perring, and F. Do\u{g}an, Nature {\bf 395 }, 580 (1998).

%\bibitem{Mook-Nature}  H.A. Mook {\em et al.},  Nature {\bf 404}, 729 (2000).

% \bibitem{Bourges-Miami}  P. Bourges {\em et al.},   
% in {\it High Temperature Superconductivity } S.E. Barnes {\it et al} 
% Eds. (CP483 AIP, Amsterdam, 1999)  pp 207-212 (cond-mat/9902067).

\bibitem{Arai}  M. Arai, T. Nishijima, Y. Endoh, T. Egami, S. Tajima, K. Tomimoto, {\em et al.}, Phys. Rev. Lett. {\bf 83}, 608 (1999).

\bibitem{Bourges-science}  % P. Bourges {\em et al.}, 
P. Bourges, Y. Sidis, H. F. Fong, L. P. Regnault, J. Bossy, A. Ivanov, and B. Keimer, Science {\bf 288}, 1234 (2000).

% ---------------------------------- YBCO- incom

\bibitem{dmitry} D. Reznik, P. Bourges, L. Pintschovius, Y. Endoh, {\em et al.}, submitted to Phys. Rev. Lett
(cond-mat/0307591).

\bibitem{buyers} C. Stock, W. J. L. Buyers,  R. Liang,  D. Peets,  Z. Tun,  {\em et al.}, 
Phys. Rev. B, (2004) (cond-mat/0308168).

\bibitem{pailhes} S. Pailh\`es, Y. Sidis, P. Bourges, C. Ulrich, {\em et al.},
Phys. Rev. Lett. {\bf 91}, 237002 (2003) (cond-mat/0308394).

\bibitem{morr} D.K. Morr and D. Pines, Phys. Rev. Lett. {\bf 81}, 1086
(1998).

\bibitem{sachdev} S. Sachdev, C. Buragohain, and M. Vojta, Science {\bf 286}, 2479 (1999).

% \bibitem{Onufrieva99} F. Onufrieva, and P. Pfeuty, cond-mat/9903097.

\bibitem{Liu95} D. Z. Liu, Y. Zha, and K. Levin, Phys. Rev. Lett. {\bf 75}, 4130 (1995).

\bibitem{Millis96} A.J. Millis and H. Monien, Phys. Rev. B {\bf 54}, 16172 (1996).

\bibitem{Abanov99} A. Abanov, and A. V. Chubukov, Phys. Rev. lett. {\bf 83}, 1652 (1999).

\bibitem{Norman00} M.N. Norman, Phys. Rev. B {\bf 61}, 14751 (2000).

\bibitem{eremin} D. Manske, I. Eremin and K.H. Bennemann, Phys. Rev. B {\bf 63}, 054517 (2001).

\bibitem{Onufrieva02} F. Onufrieva, and P. Pfeuty, Phys. Rev. B {\bf 65}, 014502 (2002)
(cond-mat/9903097).

\bibitem{bl} J. Brinckmann and P.A. Lee, Phys. Rev. B {\bf 65}, 014502 (2002).

\bibitem{sega} I. Sega,  P. Prelov\u{s}ek, and J. Bon\u{c}a, Phys. Rev. B {\bf 68}, 054524 (2003).

\bibitem{raman} For a recent review, see e.g. M. Cardona, Physica C
{\bf 318}, 30 (1999).

\bibitem{fermisurface} H. Ding, M. R. Norman, T. Yokoya, T. Takeuchi, M. Randeria,
 {\it et al.}, Phys. Rev. Lett. {\bf 78}, 2628 (1997).

\bibitem{Li99} T. Li and Z. Gan, Phys. Rev B {\bf 60}, 3092 (1999); T. Li, Phys. Rev. B {\bf 64}, 012503 (2001).

\bibitem{Reznik96} D. Reznik, P. Bourges, H. F. Fong, L. P. Regnault, J. Bossy 
       C. Vettier, {\em et al.}, Phys. Rev. B {\bf 53} R14741 (1996).

% \bibitem{Bourges-LT23} P. Bourges {\em et al.}, to appear in LT23 Proceedings.
% ---------------------------------- SO(5)

\bibitem{Zhang97} S.C. Zhang, Science {\bf 275}, 1089 (1997).

\bibitem{Demler98}  E. Demler, H. Kohno, and S.C. Zhang,  Phys. Rev. B {\bf 58}, 5719 (1998).

\bibitem{Hu01} J.P. Hu and S.C. Zhang, Phys. Rev. B {\bf 64}, 100502 (2001).

\bibitem{Chubukov-pi} O. Tchernychyov, M. R. Norman and A. V. Chubukov, Phys. Rev. B {\bf 63}, 144507 (2001).

% ---------------------------------- stripes

% \bibitem{Emery-review} E. W. Carlson {\em et al.}, Review chapter to appear in {\it The Physics of Conventional and Unconventional Superconductors},  ed. by K. H. Bennemann and J. B. Ketterson (Springer-Verlag), cond-mat/0206217. 

\bibitem{Batista} C.D. Bastista, G. Ortiz, and A.V. Balatsky, Phys. Rev. B {\bf 65}, 180402 (2002).

\bibitem{kruger} F. Kr\"uger, and S. Scheidl, Phys. Rev. B {\bf 67}, 134512  (2003).

\bibitem{Morais-smith}  % . Hasselmann {\em et al.}, 
N. Hasselmann, A. H. Castro Neto, C. Morais Smith, and Y. Dimashko, Phys. Rev. lett. {\bf 82}, 2135 (1999).

\bibitem{lsno} P. Bourges, Y. Sidis, M. Braden, K. Nakajima, and J.M. Tranquada,
Phys. Rev. Lett., {\bf 90}, 147202 (2003).

\end{thebibliography}
\end{document}